\documentclass[sigconf,final,authorversion]{acmart}

\AtBeginDocument{%
  \providecommand\BibTeX{{%
    \normalfont B\kern-0.5em{\scshape i\kern-0.25em b}\kern-0.8em\TeX}}}

\copyrightyear{2022}
\acmYear{2022}
\setcopyright{acmlicensed}
\acmConference[MSR '22]{19th International Conference on Mining Software Repositories}{May 23--24, 2022}{Pittsburgh, PA, USA}
\acmBooktitle{19th International Conference on Mining Software Repositories (MSR '22), May 23--24, 2022, Pittsburgh, PA, USA}
\acmPrice{15.00}
\acmDOI{10.1145/3524842.3528503}
\acmISBN{978-1-4503-9303-4/22/05}

\acmSubmissionID{83}

\usepackage[binary-units]{siunitx}
\usepackage{listings}
\usepackage{float}

\newfloat{lstfloat}{htbp}{lop}
\floatname{lstfloat}{Listing}

\usepackage[inline]{enumitem}
\usepackage[acronym]{glossaries}
\newglossaryentry{repominer}
{
        name=pyrepositoryminer,
        description={}
}
\newacronym{msr}{MSR}{Mining Software Repositories}
\newacronym{vcs}{VCS}{version control systems}

\definecolor{codegreen}{rgb}{0,0.6,0}
\definecolor{codegray}{rgb}{0.5,0.5,0.5}
\definecolor{codepurple}{rgb}{0.58,0,0.82}

\lstdefinestyle{basicstyle}{
    captionpos=b,
    numbers=left,
    numberstyle=\small,
    numbersep=8pt,
    frame = single
}

\lstdefinestyle{highlightstyle}{
    commentstyle=\color{codegreen},
    keywordstyle=\color{magenta},
    numberstyle=\tiny\color{codegray},
    stringstyle=\color{codepurple},
    captionpos=b,
    numbers=left,
    numberstyle=\small,
    numbersep=8pt,
    frame = single
}

\usepackage{booktabs}

\usepackage{tikz}
\usetikzlibrary{calc,fit,backgrounds,positioning,shapes,snakes,arrows.meta,spy,shapes.multipart,patterns}

\usepackage{pgfplots}
\usepackage{pgfplotstable}
\usepackage{pgfplotstablefilter}
\pgfplotsset{compat=newest}
\usepgfplotslibrary{statistics}

\definecolor{plots-1}{RGB}{228,26,28}
\definecolor{plots-2}{RGB}{55,126,184}
\definecolor{plots-3}{RGB}{77,175,74}
\definecolor{plots-4}{RGB}{152,78,163}
\definecolor{plots-5}{RGB}{255,127,0}
\definecolor{plots-6}{RGB}{153,153,153}
\definecolor{plots-7}{RGB}{166,86,40}

\usepackage{csquotes}

\makeatletter
\newcommand{\showfont}{encoding: \f@encoding{},
  family: \f@family{},
  series: \f@series{},
  shape: \f@shape{},
  size: \f@size{}
}
\newcommand{\iffont}[3]{\ifthenelse{\equal{\f@family}{#1}}{#2}{#3}}
\makeatother

\begin{document}

\title{Tooling for Time- and Space-efficient git Repository Mining}

\author{Fabian Heseding}
\email{fabian.heseding@student.hpi.uni-potsdam.de}
\affiliation{%
  \institution{Hasso Plattner Institute, Digital Engineering Faculty, University of Potsdam}
  \streetaddress{Prof.-Dr.-Helmert-Str. 2--3}
  \city{Potsdam}
  \country{Germany}
  \postcode{14482}
}
\author{Willy Scheibel}
\email{willy.scheibel@hpi.uni-potsdam.de}
\orcid{0000-0002-7885-9857}
\affiliation{%
  \institution{Hasso Plattner Institute, Digital Engineering Faculty, University of Potsdam}
  \streetaddress{Prof.-Dr.-Helmert-Str. 2--3}
  \city{Potsdam}
  \country{Germany}
  \postcode{14482}
}
\author{J{\"u}rgen D{\"o}llner}
\email{juergen.doellner@hpi.uni-potsdam.de}
\affiliation{%
  \institution{Hasso Plattner Institute, Digital Engineering Faculty, University of Potsdam}
  \streetaddress{Prof.-Dr.-Helmert-Str. 2--3}
  \city{Potsdam}
  \country{Germany}
  \postcode{14482}
}

\renewcommand{\shortauthors}{Heseding, Scheibel, and D{\"o}llner}

\begin{abstract}
Software projects under version control grow with each commit, accumulating up to hundreds of thousands of commits per repository.
Especially for such large projects, the traversal of a repository and data extraction for static source code analysis poses a trade-off between granularity and speed.

We showcase the command-line tool \gls*{repominer} that combines a set of optimization approaches for efficient traversal and data extraction from git repositories while being adaptable to third-party and custom software metrics and data extractions.
The tool is written in Python and combines bare repository access, in-memory storage, parallelization, caching, change-based analysis, and optimized communication between the traversal and custom data extraction components.
The tool allows for both metrics written in Python and external programs for data extraction.
A single-thread performance evaluation based on a basic mining use case shows a mean speedup of $15.6\times$ to other freely available tools across four mid-sized open source projects.
A multi-threaded execution allows for load distribution among cores and, thus, a mean speedup up to $86.9\times$ using 12 threads.

\end{abstract}




\maketitle


\section{Introduction}


With software playing a key role in the modern world, gaining insights about software becomes crucial~\cite{ma2019world}.
Such insights entail quality metrics about the software as well as data about the development process.
One approach to gaining insights about software, especially its source code, is deriving it from their repositories that are governed by \acrfull{vcs} such as git~\cite{bird2009promises}.
\acrfull*{msr} focuses on extracting and analyzing data available in software repositories to uncover interesting, helpful, and actionable information about the software system~\cite{spadini2018pydriller}.
One goal is to quantify and monitor the quality of the developed software employing software metrics~\cite{limberger2019-softwaremap}, another is to explore new ways of making sense of the data.
However, traversing and gathering the vast amount of data poses a challenge~\cite{mattis2020three,ma2019world,ma2021world,dyer2013boa}.

\begin{figure}[t]
    \centering
    \scriptsize
    \begin{tikzpicture}[inner sep=0pt,outer sep=0pt,
        my triangle/.style={-{Triangle[width=\the\dimexpr3.0\pgflinewidth,length=\the\dimexpr3.0\pgflinewidth]}}
        ]
        \node[rectangle,text=black,draw=black!20,rounded corners=2pt,minimum width=1.5cm,minimum height=0.8cm,text width=1.45cm,align=center] (software-repository) at (0.0cm, 0.0cm) {{Software\\Repository}};
        \node[rectangle,text=black,draw=black!20,rounded corners=2pt,minimum width=1.5cm,minimum height=0.8cm,text width=1.45cm,align=center] (file-provisioning) at (2.3cm, -0.7cm) (file-provisioning) {{File\\Provisioning}};
        \node[rectangle,text=black,draw=black!20,rounded corners=2pt,minimum width=1.5cm,minimum height=0.8cm,text width=1.45cm,align=center] (metrics-calculation) at (4.6cm, -0.7cm) {{Metrics\\Calculation}};
        \node[rectangle,text=black,draw=black!20,rounded corners=2pt,minimum width=1.5cm,minimum height=0.8cm,text width=1.45cm,align=center] (each-branch-commit) at (3.45cm, 0.7cm) {{Each branch\\Each commit}};
        \node[rectangle,text=black,draw=black!20,rounded corners=2pt,minimum width=1.5cm,minimum height=0.8cm,text width=1.45cm,align=center] (merge-outputs) at (6.9cm, 0.0cm) {{Merge\\Outputs}};
        \node[rectangle,text=black,draw=black!20,rounded corners=2pt,minimum width=1.5cm,minimum height=0.8cm,text width=1.45cm,align=center] (analysis-reports-visualization) at (6.9cm, 1.4cm) {{Analysis \&\\Reports \&\\Visualization}};
        \node[below=0.2cm of software-repository] { \texttt{\$ git clone} };
        \node[below=0.2cm of file-provisioning] { \texttt{\$ git checkout} };
        \node[below=0.2cm of metrics-calculation,text width=2.4cm,align=center] { \texttt{\$ ./script\_A.sh . \&\&\\radon *.py \&\& ...} };
        \node[above=0.2cm of each-branch-commit] { \texttt{for commit in commits ...} };
        \node[above=0.2cm of analysis-reports-visualization] { \texttt{\$ jupyter lab} };
        \draw[draw=black!20,line width=1.5pt,my triangle,shorten <= 3pt,shorten >= 3pt] (software-repository.east) -- +(0.8cm,0cm);
        \draw[draw=black!20,line width=1.5pt,my triangle,shorten <= 3pt,shorten >= 3pt] (file-provisioning.east) -- (metrics-calculation);
        \draw[draw=black!20,line width=1.5pt,my triangle,shorten <= 3pt,shorten >= 3pt] (metrics-calculation.north) -- (each-branch-commit.south east);
        \draw[draw=black!20,line width=1.5pt,my triangle,shorten <= 3pt,shorten >= 3pt] (each-branch-commit.south west) -- (file-provisioning.north);
        \draw[draw=black!20,line width=1.5pt,my triangle,shorten <= 3pt,shorten >= 3pt] (merge-outputs.west) +(-0.8cm,0cm) -- (merge-outputs.west);
        \draw[draw=black!20,line width=1.5pt,my triangle,shorten <= 3pt,shorten >= 3pt] (merge-outputs.north) -- (analysis-reports-visualization.south);
    \end{tikzpicture}
    \caption{Example process for software analysis that requires repository mining across multiple branches and commits, while computing multiple metrics using third-party and custom tools. The results can than be used for further analysis, static reports, or interactive visualization.}
    \label{fig:process}
\end{figure}
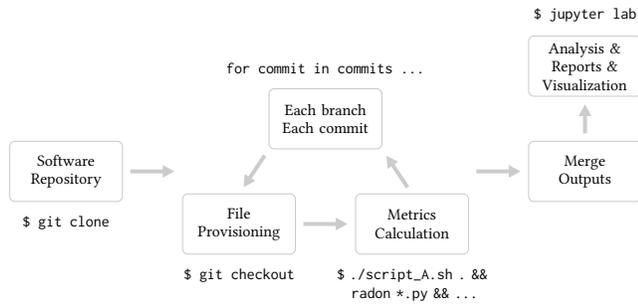

For an exemplary analysis of a software repository, we assume a process that covers the following parts:
\begin{enumerate*}
    \item Acquiring the software repository (e.g., \texttt{git clone}),
    \item Provisioning the files to run an analysis on (e.g., \texttt{git checkout}),
    \item Running an analysis on the files (applying existing tools or custom scripts that derive quality metrics), and
    \item Reporting the results (e.g., by creating reports or interactive visualization)
\end{enumerate*}.
During this process, the analysis usually covers parts or the whole repository and, thus, requires some sort of traversal (Figure~\ref{fig:process}).
With a list of revisions to analyze, each revision is checked out to the working directory.
Upon each checkout, tools that calculate software metrics are run and the results are logged.
After handling all revisions, the results are merged and reported.
%
For example, a linter may run on a working directory and output a quality score of the source code.
Another example is to measure the source code using software metrics that report higher-level information.
Tracking those scores across versions allows for detecting changes, trends, and insights about a software project's development status and quality.

If an applied software developments process includes regular software analysis, the data mining from the repository gets demanding over time.
This demand occurs through
\begin{enumerate*}
    \item ever-growing source code,
    \item the number of changes created during the overall development,
    \item the number of concurrent development branches, e.g., through multiple developers or multiple maintained versions,
    \item the complexity of the analyses performed, and
    \item the expectation for timely results for effective use in the development process.
\end{enumerate*}




In this paper, we showcase a low-level analysis tool for efficient git repository mining named \gls*{repominer}~\cite{heseding2022-pyrepositoryminer} and discuss architectural and design decisions that improve the execution speed of the tool.
It is written in Python, available at PyPi, and serves as a framework that provides repository traversal and allows to use custom metrics provided as scripts and integrate existing metric calculation tools.
A first analysis using the mid-sized open-source projects numpy, matplotlib, pandas, and tensorflow shows greatly improved execution times compared to other low-level source code repository \acrshort*{msr} tools.
Although provisioned as a separate tool focusing on Python and git, the discussed approaches are applicable to other \acrshort*{msr} tools and \acrshort*{vcs}, too.



\section{Related Work}\label{sec:relatedwork}

In the last years, many tools for mining git and related systems were proposed and researched.
Although a number \acrshort*{vcs} are suitable for mining of software data, the specific ecosystem around git seems to be a focus.
As such, git, a decentralized version control system, and Github, the software project management platform built around git repositories, are driving best practices in software development, software analyses, and repository mining~\cite{bird2009promises,kalliamvakou2014promises}.
When faced with an \acrshort*{msr} task, researchers and practitioners have the choice to use one of the few existing tools and frameworks that focus on mining, use a higher-level system or infrastructure where they integrate their analysis in, or develop their own tool tailored to their specific needs.
Existing tools and infrastructure, however, often require learning a domain-specific language to formulate the analysis in or constrain the analysis to a predefined use case.

\paragraph{\acrshort*{msr} Tools.}
A tool similar in usage to our implementation is PyDriller\footnote{\url{https://github.com/ishepard/pydriller}}~\cite{spadini2018pydriller}, an open-source Python framework for Mining Software Repositories.
PyDriller is a high-level interface around the comparatively low-level GitPython\footnote{\url{https://github.com/gitpython-developers/GitPython}} library that offers ease of use for everyday \acrshort*{msr} tasks with minimal to no decrease in performance.
While PyDriller offers an easy to learn interface for Python developers, its performance is heavily dependent on the implementation of the concrete task at hand.
For example, when iterating over all of the blobs of all commits and calculating metrics, unchanged files are not skipped even though the metrics are already calculated.
Additionally, this example would run on one core if the researcher does not implement any parallelization.

Another tool that aims to simplify mining tasks is RepoFS\footnote{\url{https://github.com/AUEB-BALab/RepoFS}}~\cite{salis2019repofs}, a virtual filesystem mapping to a Git repository.
RepoFS offers directories for each revision within the filesystem and maps
access to a bare repository.
Thus, it enables to run filesystem-based mining tools across revisions without incurring disk IO from repeated checkouts.
However, it leaves parallelization and efficient calculation logic to the actual mining tool or task implementation.

With regards to mining tools dealing primarily with GitHub, we found tools for efficiently mining artifacts from GitHub, such as ModelMine~\cite{reza2020modelmine}, as well as for efficient classification of commits for downstream analysis, such as GitCProc~\cite{casalnuovo2017gitcproc}, using primarily regular expressions.
While these tools excel in their target use cases, they are not suited for general purpose analysis of a local repository.

\paragraph{\acrshort*{msr} Infrastructure.}
A prominent tool in the \acrshort*{msr} research field tool is Boa~\cite{dyer2013boa}, a domain-specific language and infrastructure for \acrshort*{msr}.
Boa enables researchers to reproduce data extraction and analysis by specifying mining tasks in the Boa language and running them on the Boa infrastructure.
The Boa infrastructure also offers scalability for expensive tasks.
Boa, however, does not focus on efficiency in calculations.
Additionally, researchers need to learn a new domain-specific language compared to a few shell commands.
Finally, flexibility in custom metric logic is limited by what the language allows.

Other tools offering an infrastructure framework include Crossflow\footnote{\url{https://github.com/crossflowlabs/crossflow}}~\cite{kolovos2019crossflow}, SmartSHARK\footnote{\url{https://github.com/smartshark/smartshark.github.io}}~\cite{trautsch2020smartshark}, and World~of~Code~\cite{ma2019world,ma2021world}.
CrossFlow is a domain specific language and framework that offers scalability for \acrshort*{msr} tasks; SmartSHARK aggregates data from different sources in a harmonized schema.
The World~of~Code is an infrastructure that enables research on free and open-source software repositories.
While these tools offer scalability, they require researchers to learn a new framework and run their analyses on their infrastructure.
This requirement restricts the flexibility of these tools for general-purpose analysis.
Finally, projects such as the Three~Trillion~Lines~\cite{mattis2020three} describe the required infrastructure setup to manage large-scale software analysis.
While fault tolerance and domain-specific data structures are important to a large-scale \acrshort*{msr} endeavor, we describe an approach to make the actual mining of repositories more efficient.

\section{Approaches}\label{sec:approach}


Efficiency in repository mining can be described by executing only required operations, omitting as much as possible intermediate operations that do not directly contribute to the results, as well as leveraging multi-threading capabilities of current hardware.

\paragraph{RAM Disk.}
As such, physical representation of files on a disk is an intermediate artifact that can be replaced with a RAM Disk that would improve read and write access during traversal of a repository.
Eventual writes during mining would also pose no additional strain on the lifetime of HDDs and SSDs.
As creation, management, and usage of a RAM Disk is usually done using the operating system and not on a tool-level, this approach is feasible for all \acrshort*{msr} tools.

\paragraph{Bare Repository Access.}
Further, checking out revisions into a working directory is inefficient as it incurs disk IO by reading the files from the repository folder and writing them to the working directory~\cite{salis2019repofs}.
Similarly, a working directory creates a copy of data already available in a git repository, which can be avoided if analyses and tools are versatile to handle input data using in-memory interfaces.
With bare repository access -- a standard concept of git -- the used memory on disk is strictly bound to the space requirements of the bare repository.
Additionally, using a working directory complicates parallelization, i.e., a parallel computation would require one working directory per thread.

\paragraph{Parallelization.}
We apply concurrent computation of metrics and source code analyses on a per-commit level.
Such parallelization is feasible if the results for different commits can be computed independently.
As repository mining is a task that is inherently idempotent and should only require read access to a software repository, this holds on a conceptual level.
From an implementation perspective, this also holds if working directories can be omitted, or, alternatively, each thread uses its own working directory.

\paragraph{Caching.}
Regarding metric calculations, the idempotent property of analyses allow for thorough caching of results.
This is a large advantage for software repositories, as each commit usually changes only parts of the source code~\cite{arafat2009-commitsizes}.
If, for example, a blob rarely changes across revisions but is analyzed for each revision nonetheless, these calculations are redundant and should be cached instead.

\paragraph{Change-based Analysis.}
Likewise, some implementations of analyses or metrics operate on a whole working directory instead of only changed files for a commit. However, unchanged files were presumably mined with an earlier commit during the analysis.
This optimization can be leveraged using a cache and filtering for files that were not mined in an earlier commit. Even external tools can profit from this filter.

\paragraph{Optimized Communication between Components.}
Last, we consider an optimized communication through in-memory string representation of files.
This optimization aims to reduce file operations and relate to the access of the bare repository and handling of the repository on a RAM disk.
The communication between the traversing component and analysis components is ideally done through contents residing in RAM and not on disk.


\section{Implementation}\label{sec:implementation}

\begin{figure}
    \centering
    \includegraphics[width=1.0\linewidth]{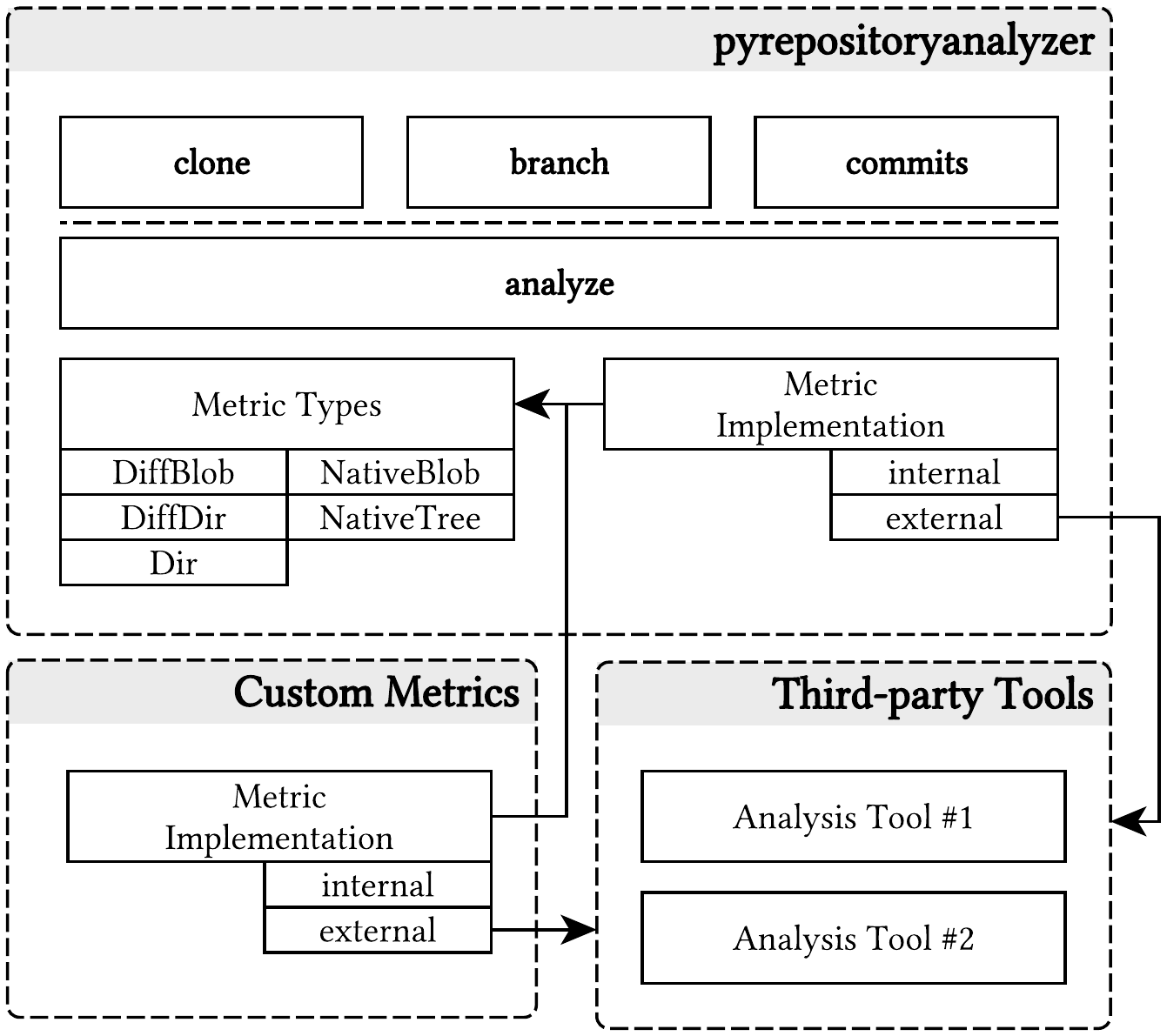}
    \caption{Architecture of the \gls*{repominer}. Custom metrics can be developed as (1) internal metrics using standard Python libraries and source code, or (2) external metrics by wrapping third-party tools, both implemented by subclassing one of the metric types in Python scripts. The custom metrics can be developed, maintained and stored independently from the \gls*{repominer}.}
    \label{fig:architecture}
\end{figure}

We target all the performance optimizations with our tool and further focus on usability, extensibility, and integration.
As a result, we propose a tool that applies the optimization approaches internally -- by architecture and design -- and that allows further extensions with user-specific analyses and metrics (\autoref{fig:architecture}).
We implemented the tool as an open-source Python project available on PyPi\footnote{\href{https://pypi.org/project/pyrepositoryminer/}{https://pypi.org/project/pyrepositoryminer/}} with its source code hosted on GitHub\footnote{\url{https://github.com/fabianhe/pyrepositoryminer}}, published under the GPL-3.0 License.
We leverage pygit2\footnote{\url{https://github.com/libgit2/pygit2}}, the python bindings for libgit2\footnote{\url{https://github.com/libgit2/libgit2}}, for direct interaction with Git objects of the bare repository.
The command-line interface is built using the Typer\footnote{\url{https://github.com/tiangolo/typer}} framework and comprises four commands:
\begin{enumerate*}
    \item \texttt{clone},
    \item \texttt{branch},
    \item \texttt{commits}, and
    \item \texttt{analyze}
\end{enumerate*}.

\paragraph{Command-line Interface.}
The \texttt{clone} command clones a bare repository from a URL pointing to a git repository into a local path.
The \texttt{branch} command returns the branches of a repository. Remote and local branches can be included or excluded using the command options.
The \texttt{commits} command gets the commit hashes of commits on the input branches.
By default, merge commits are simplified to their first parent commit, duplicates are eliminated, and the commits are sorted in topological order.
This behavior can be changed using command options.
Additionally, the output can be limited to the first $n$ commits encountered.
The \texttt{analyze} command analyzes all input commits.
Arguments to the analyze command are names of the predefined metrics to use.
This tool further supports command options, e.g., to load implementations of custom metrics by using the \texttt{{-}-custom-metrics} option, or to configure the number of worker processes for parallel analysis (\texttt{{-}-workers}).

\paragraph{Tool Operation.}
As an example, we demonstrate how to analyze the numpy repository hosted on GitHub with regards to the Halstead metrics and the total lines of code on all commits of the main branch (see also \autoref{lst:example}):
This example \begin{enumerate*}
    \item mounts a RAM disk of size \SI{250}{\mega\byte},
    \item clone the numpy repository from GitHub to the newly created RAM disk using the \texttt{clone} command,
    \item collects local branches,
    \item and their reachable commits -- topologically sorted --, and
    \item perform computation of the Halstead and Lines-of-Code metrics on each commit.
\end{enumerate*}.
The results are piped to a file for reporting.
As cleanup, and the RAM disk is unmounted.

\begin{lstfloat}[t]
\begin{lstlisting}[language=bash, caption={Running the \gls*{repominer} on macOS. The content of the ramdisk scripts are available in the repository of the tool.},  basicstyle=\footnotesize, label=lst:example]
> ./mount-ramdisk-macos.sh 512000
> pyrepositoryminer clone https://github.com/numpy/numpy \
  /Volumes/RAMDisk/numpy.git
> pyrepositoryminer branch /Volumes/RAMDisk/numpy.git \
  | pyrepositoryminer commits /Volumes/RAMDisk/numpy.git \
  | pyrepositoryminer analyze /Volumes/RAMDisk/numpy.git \
  halstead loc > out.jsonl
> ./unmount-ramdisk-macos.sh disk3
\end{lstlisting}
\end{lstfloat}

\paragraph{Metric Types.}
Metrics are implemented as subclasses of their respective metric type.
Each superclass corresponds to the input format and intermediate calculations that can be shared across all metrics of that metric type.
We provide classes for file blobs (\texttt{NativeBlobMetric}), commit trees (\texttt{TreeMetric}), and directories (\texttt{DirMetric}), each with the option of only using the touched blobs of a commit (\texttt{DiffBlobMetric}, \texttt{DiffTreeMetric}, \texttt{DiffDirMetric}).
We encourage to write metrics that use native git objects rather than a working directory for analysis.
Metrics that request a checked-out working directory, i.e., subclasses of the \texttt{DirMetric} and the \texttt{DiffDirMetric}, are handled by providing a working directory nonetheless.
%
For example, the class of a directory of touched blobs - \texttt{DiffDirMetric} - finds all touched blobs of a revision and checks out a working directory on which filesystem-based tools can run.


\paragraph{User-defined Metrics.}
Custom metrics can be easily implemented by using the library interface of the tool.
The tool provides five superclasses to subtype from -- each corresponding to one of the supported metric types -- such that own analyses and metrics can be implemented in scripts and get loaded together with the tool.
Existing tools can be integrated by implementing them as a metric.
For example, an executable of a static source code analysis tool such as Tokei\footnote{\href{https://github.com/XAMPPRocky/tokei}{https://github.com/XAMPPRocky/tokei}} can get wrapped as a metric by using the \texttt{DiffDir} metric and inter-process communication and string conversion.

\section{Evaluation}\label{sec:evaluation}
We evaluate the \gls*{repominer} using two performance measurements:
\begin{enumerate*}
    \item the time required to iterate all commits of a repository and
    \item the runtime of a basic mining use case
\end{enumerate*}.
The measurements are compared to similar implementations using PyDriller.
We measured on an Intel Core i9 CPU with 18 cores at 3.00\,GHz and 128\,GB of main memory on an Ubuntu 20.04.
The software repositories used for the measurements are four publicly available mid-sized GitHub repositories (Table~\ref{tab:projects}).
Before measuring either tool, we loaded the repositories onto a RAM disk.

\begin{figure}[t]
    \centering%
    \pgfplotstableread[col sep=comma,]{measurements/traversal-performance.csv}\datatable%
    \pgfplotstablefilter[implementation equals {PyDriller}]{\datatable}{\datatablePyDriller}%
    \pgfplotstablefilter[implementation equals {pyrepositoryminer}]{\datatable}{\datatablePyrepositoryminer}%
    \begin{tikzpicture}[outer sep=0pt]%
        \begin{axis}[
        	compat=newest,
        	name=traversal,
        	ybar=5pt,
        	axis on top,
        	scaled y ticks=false,
        	width=0.75\columnwidth,
        	height=12\baselineskip,
        	enlarge x limits=0.15,
        	bar width=6.0pt,
        	ylabel=\footnotesize{commits per second},
        	xlabel=\footnotesize{Project},
        	ymajorgrids,
        	major grid style={draw=white},
        	tick label style={font=\scriptsize},
        	yticklabel style={
        		/pgf/number format/fixed,
        		/pgf/number format/precision=0
        	},
        	axis x line*=bottom,
        	axis y line*=left,
        	ytick align=outside,
        	xtick=data,
        	xticklabels from table={\datatable}{project},
        	ymin=0.0,
        	ymax=50000,
        	point meta=rawy,
        	nodes near coords,
        	every node near coord/.append style={rotate=90, anchor=west, font=\scriptsize,/pgf/number format/.cd, precision=0, fixed,1000 sep={\,}},
	        nodes near coords align={vertical},
    	]
        	\addplot[white,very thin,fill=plots-1,text=black] table[y={commits_per_second},x expr=\coordindex]{\datatablePyDriller};
        	\addplot[white,very thin,fill=plots-2,text=black] table[y={commits_per_second},x expr=\coordindex]{\datatablePyrepositoryminer};
    	\end{axis}
    	\begin{axis}[
        	name=traversal_legend,
        	compat=1.3,
        	at={($(traversal.north east)+(0.0cm,0.0cm)$)},
        	hide axis,
        	bar width=0pt,
        	ticks=none,
        	legend style={
        		at={(0.0,0.0)},
        		anchor=north west,
        		draw=none,
        		legend cell align=left,
        	},
        	legend image code/.code={%
        		\draw (0.0cm,-0.05cm) rectangle (0.4cm,0.15cm);
        	},
        	text height=1.1ex,
        	height=2cm,
        	ymin=0.0,
        	ymax=1.0,
        	xmin=0.0,
        	xmax=1.0
        ]
        	\addplot[draw=none,white,very thin,fill=plots-1] coordinates {(0,0)};
        	\addlegendentry{\;\scriptsize PyDriller}
        	\addplot[draw=none,white,very thin,fill=plots-2] coordinates {(0,0)};
        	\addlegendentry{\;\scriptsize \gls*{repominer}}
        \end{axis}%
    \end{tikzpicture}%
    \vspace{-0.5\baselineskip}
    \caption{Throughput of commit iteration on all reachable commits on all remote branches.}
    \label{fig:tp2}
\end{figure}
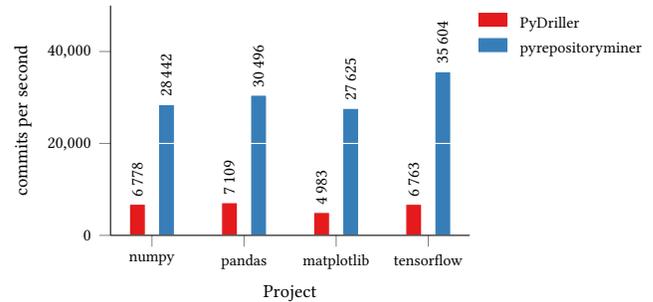

\paragraph{Commit Iteration Throughput.}
In our commit iteration use case, the task is to get and print each revision id on a repository's main branch.
We aim to measure the framework overhead cost without a specific software metric.
Using the \gls*{repominer}, this is accomplished by running and piping the inputs and outputs of the \texttt{branch} and \texttt{commits} commands without any further configuration.
Using PyDriller, this is accomplished by running the \texttt{traverse\_commits} method.
%
As a result, the \gls*{repominer} achieves a minimum throughput of \SI{23229}{commits\per\second} and a maximum of \SI{28899}{commits\per\second}, whereas PyDriller achieves a throughput between \SI{4191}{commits\per\second} and \SI{6155}{commits\per\second} (see Figure~\ref{fig:tp2}).
The mean speedup achieved by the \gls*{repominer} compared to PyDriller is at $4.8\times$.
When analyzing large software repositories with more than \SI{60000}{commits}, the total runtime of iterating commits stays within a few seconds.

\begin{table}[t]
    \pgfplotstableread[col sep=comma,]{measurements/commit_counts.csv}\commitsTable%
    \pgfplotstablefilter[project equals {numpy}]{\commitsTable}{\commitsTableNumpy}%
    \pgfplotstablegetelem{0}{number_of_commits}\of\commitsTableNumpy
    \pgfmathsetmacro{\numberOfCommitsNumpy}{\pgfplotsretval}
    \pgfplotstablefilter[project equals {pandas}]{\commitsTable}{\commitsTablePandas}%
    \pgfplotstablegetelem{0}{number_of_commits}\of\commitsTablePandas
    \pgfmathsetmacro{\numberOfCommitsPandas}{\pgfplotsretval}
    \pgfplotstablefilter[project equals {matplotlib}]{\commitsTable}{\commitsTableMatplotlib}%
    \pgfplotstablegetelem{0}{number_of_commits}\of\commitsTableMatplotlib
    \pgfmathsetmacro{\numberOfCommitsMatplotlib}{\pgfplotsretval}
    \pgfplotstablefilter[project equals {tensorflow}]{\commitsTable}{\commitsTableTensorflow}%
    \pgfplotstablegetelem{0}{number_of_commits}\of\commitsTableTensorflow
    \pgfmathsetmacro{\numberOfCommitsTensorflow}{\pgfplotsretval}
    \centering
    \caption{Open source projects used for evaluation. The reported number of commits include all commits reachable by all remote branches.}
    \label{tab:projects}
    \vspace{-0.75\baselineskip}
    \small
	\setlength{\tabcolsep}{4.0pt}%
    \begin{tabular}{l|l|l}
        \toprule
        \textbf{Identifier} & \textbf{Repository} & \textbf{\# Commits}\\ \midrule
        numpy & \href{https://github.com/numpy/numpy.git}{https://github.com/numpy/numpy.git} & \num{\numberOfCommitsNumpy} \\
        matplotlib & \href{https://github.com/matplotlib/matplotlib.git}{https://github.com/matplotlib/matplotlib.git} & \num{\numberOfCommitsMatplotlib} \\
        pandas & \href{https://github.com/pandas-dev/pandas.git}{https://github.com/pandas-dev/pandas.git} & \num{\numberOfCommitsPandas} \\
        tensorflow & \href{https://github.com/tensorflow/tensorflow.git}{https://github.com/tensorflow/tensorflow.git} & \num{\numberOfCommitsTensorflow} \\
        \bottomrule
    \end{tabular}
\end{table}

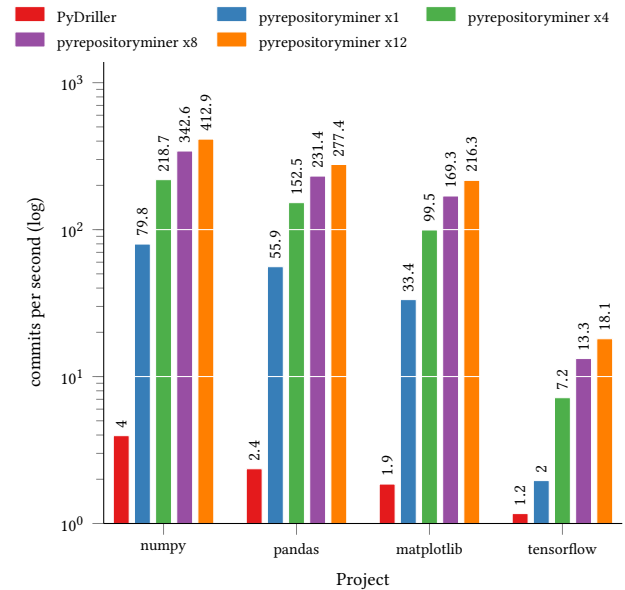
\begin{figure}[t]
    \centering%
    \pgfplotstableread[col sep=comma,]{measurements/mining-performance.csv}\datatable%
    \pgfplotstablefilter[implementation equals {PyDriller}]{\datatable}{\datatablePyDriller}%
    \pgfplotstablefilter[implementation equals {pyrepositoryminer1}]{\datatable}{\datatablePyrepositoryminerOne}%
    \pgfplotstablefilter[implementation equals {pyrepositoryminer4}]{\datatable}{\datatablePyrepositoryminerFour}%
    \pgfplotstablefilter[implementation equals {pyrepositoryminer8}]{\datatable}{\datatablePyrepositoryminerEight}%
    \pgfplotstablefilter[implementation equals {pyrepositoryminer12}]{\datatable}{\datatablePyrepositoryminerTwelve}%
    \begin{tikzpicture}[outer sep=0pt]%
        \begin{axis}[
        	compat=newest,
        	name=traversal,
        	ybar=2pt,
        	ymode=log,
        	axis on top,
        	enlarge x limits=0.15,
        	width=1.00\columnwidth,
        	height=20\baselineskip,
        	bar width=6.0pt,
        	ylabel=\footnotesize{commits per second (log)},
        	xlabel=\footnotesize{Project},
        	ymajorgrids,
        	major grid style={draw=white},
        	tick label style={font=\scriptsize},
        	axis x line*=bottom,
        	axis y line*=left,
        	ytick align=outside,
        	xtick=data,
        	xticklabels from table={\datatable}{project},
        	ymin=1.0,
        	ymax=1400,
        	point meta=rawy,
        	nodes near coords,
        	every node near coord/.append style={rotate=90, anchor=west, font=\scriptsize,/pgf/number format/.cd, precision=1, fixed,1000 sep={}},
	        nodes near coords align={vertical},
    	]
        	\addplot[white,very thin,fill=plots-1,text=black] table[y={commits_per_second},x expr=\coordindex]{\datatablePyDriller};
        	\addplot[white,very thin,fill=plots-2,text=black] table[y={commits_per_second},x expr=\coordindex]{\datatablePyrepositoryminerOne};
        	\addplot[white,very thin,fill=plots-3,text=black] table[y={commits_per_second},x expr=\coordindex]{\datatablePyrepositoryminerFour};
        	\addplot[white,very thin,fill=plots-4,text=black] table[y={commits_per_second},x expr=\coordindex]{\datatablePyrepositoryminerEight};
        	\addplot[white,very thin,fill=plots-5,text=black] table[y={commits_per_second},x expr=\coordindex]{\datatablePyrepositoryminerTwelve};
    	\end{axis}
    	\begin{axis}[
        	name=traversal_legend,
        	compat=1.3,
        	at={($(traversal.north east)+(0.0cm,0.0cm)$)},
        	hide axis,
        	bar width=0pt,
        	ticks=none,
        	legend columns=3,
        	legend style={
        		at={(0.0,0.0)},
        		anchor=south east,
        		draw=none,
        		legend cell align=left,
        		/tikz/column 2/.style={
                    column sep=5pt,
                },
        		/tikz/column 4/.style={
                    column sep=5pt,
                },
        	},
        	legend image code/.code={%
        		\draw (0.0cm,-0.05cm) rectangle (0.4cm,0.15cm);
        	},
        	text height=1.1ex,
        	height=2cm,
        	ymin=0.0,
        	ymax=1.0,
        	xmin=0.0,
        	xmax=1.0
        ]
        	\addplot[draw=none,white,very thin,fill=plots-1] coordinates {(0,0)};
        	\addlegendentry{\;\scriptsize PyDriller}
        	\addplot[draw=none,white,very thin,fill=plots-2] coordinates {(0,0)};
        	\addlegendentry{\;\scriptsize \gls*{repominer} x1}
        	\addplot[draw=none,white,very thin,fill=plots-3] coordinates {(0,0)};
        	\addlegendentry{\;\scriptsize \gls*{repominer} x4}
        	\addplot[draw=none,white,very thin,fill=plots-4] coordinates {(0,0)};
        	\addlegendentry{\;\scriptsize \gls*{repominer} x8}
        	\addplot[draw=none,white,very thin,fill=plots-5] coordinates {(0,0)};
        	\addlegendentry{\;\scriptsize \gls*{repominer} x12}
        \end{axis}%
    \end{tikzpicture}%
    \vspace{-0.5\baselineskip}
    \caption{Throughput of calculating all lines of code per reachable commit on all remote branches.}
    \label{fig:tp}
\end{figure}

\paragraph{Basic Mining Use Case Throughput.}
In our basic mining use case, the task is to calculate the total lines of code for each blob in each commit.
In the \gls*{repominer}, this is accomplished by implementing a blob level metric that counts each new line character in a blob's source code. 
The \gls*{repominer} ensures that they run in parallel only on modified files. 
With the PyDriller implementation, each commit is iterated in a loop and analyzed using the built-in \texttt{loc} attribute on modified files only. 
%
As a result, the \gls*{repominer} achieves a single-threaded minimum throughput of \SI{2}{commits\per\second} and a maximum of \SI{79.8}{commits\per\second} on the tested repositories, whereas PyDriller achieves a throughput between \SI{1.2}{commits\per\second} and \SI{4}{commits\per\second} (see Figure~\ref{fig:tp}).
The mean single-threaded speedup achieved by the \gls*{repominer} compared to PyDriller is at $15.6\times$, while it is $68.9\times$ as high using 12 cores.

\paragraph{Discussion.}
The results from the performance measurements seem promising.
However, iterating through a repository and providing the source code files is not the main source of computation time when applying sophisticated analysis.
Our tool aims to provide a base for fast runtimes by eliminating omittable operations and allowing for concurrent computations between commits, files, and metrics.
Additionally, our tool can be used as a wrapper to improve runtimes on existing tools, e.g., using the \texttt{DiffDir} metric, leveraging a subset of the performance optimization approaches.
The current limitations include a focus on Python for internal metrics and git as \acrshort{vcs}.
However, alternatives such as Boa~\cite{dyer2013boa} are enticing because they offer the infrastructure to perform research.
Additionally, using a custom implementation offers flexibility in research and optimization beyond what any framework approach can offer.
On the other hand, our concrete implementation suffers from the drawback that researchers need to learn our framework’s interface to use it.
Finally, further measurements need to be taken to better understand the achievable speedup compared to current tools.

\section{Conclusions}\label{sec:conclusion}

While the number of software projects and their sizes increase, and the need for software analysis and timely results remain essential, we depend on highly efficient tooling.
We approached efficiency in repository traversal and metric computation by leveraging bare repository access, in-memory storage, parallelization, caching, change-based analysis, and optimized communication between the software components.
We showcased these approaches with the tool \gls*{repominer}.
This tool is publicly hosted at Github, available at PyPi, allows the use of externally developed scripts and the integration of third-party tools for analysis
and software metrics. 
First performance results are promising and indicate a single-threaded speedup of $15.6\times$ to other freely available tools across four mid-sized open source projects and even a speedup of $86.9\times$ using 12 threads.
%
%
This allows software repository mining to be applied more frequently, thoroughly, and timely for mid-sized and large projects.
%
For future work, we target a more in-depth analysis of performance impacts on the individual approaches and a broader analysis across open source projects and industry projects.
Further, we see the potential for optimization for a change-based analysis on a line level, in contrast to the file level that is currently used.
In addition, the tool itself will profit from a broader list of available metrics and supported languages.


\begin{acks}
We want to thank the anonymous reviewers for their valuable comments and suggestions to improve this article.
This work is part of the \enquote{Software-DNA} project, which is funded by the European Regional Development Fund (ERDF or EFRE in German) and the State of Brandenburg (ILB).
This work is also part of the KMU project \enquote{KnowhowAnalyzer} (Förderkennzeichen 01IS20088B), which is funded by the German Ministry for Education and Research (Bundesministerium für Bildung und Forschung).
\end{acks}

\bibliographystyle{ACM-Reference-Format}
\bibliography{references}

\end{document}